\documentclass[12pt]{amsart}
\usepackage[english]{babel}
\usepackage{a4wide,color,graphicx,amsmath,amssymb,verbatim,mathrsfs,latexsym}

\newtheorem{theorem}{Theorem}

\newtheorem{notation}[theorem]{Notation}
\newtheorem{problem}[theorem]{Problem}
\newtheorem{proposition}[theorem]{Proposition}
\newtheorem{defi}[theorem]{Definition}
\newtheorem{lemma}[theorem]{Lemma}

\newcommand{\Set}[2]{\big\{#1\,:\,#2\big\}}

\newcommand{\R}{\mathbb{R}}		
												
\newcommand{\partd}[2]{\frac{\partial #1}{\partial #2}} 				
\newcommand{\totd}[2]{\frac{\textrm{d} #1}{\textrm{d} #2}}				
\newcommand{\tr}{\textnormal{tr}}										
\newcommand{\Tr}{\textnormal{Tr}}										
\newcommand{\Log}{\textnormal{Log}}										
\newcommand{\bk}[1]{\langle #1 \rangle}									
\newcommand{\pt}{\partial}
\newcommand{\cE}{\mathcal{E}}

\newcommand{\cS}{\mathcal{S}}
\newcommand{\cW}{\mathcal{W}}
\newcommand{\abs}[1]{\left\vert #1 \right\vert}	
\DeclareMathOperator{\Op}{Op}
\DeclareMathOperator{\sign}{sign}

\begin{document}

\title[Quantum electronic trasport in graphene]{Quantum electronic trasport in graphene: \\ a kinetic and fluid-dynamic approach}

\author{Nicola Zamponi}
\address{Dipartimento di Matematica ``U.~Dini'', Viale Morgagni 67/A, I-50134 Firenze, Italy.}
\email{nicola.zamponi@math.unifi.it}

\author{Luigi Barletti}
\address{Dipartimento di Matematica ``U.~Dini'', Viale Morgagni 67/A, I-50134 Firenze, Italy.}
\email{luigi.barletti@unifi.it}


\begin{abstract}
We derive a fluid-dynamic model for electron transport near a Dirac point in graphene.  
The derivation is based on the minimum entropy principle, which is exploited in order 
to close fluid-dynamic equations for quantum mixed states. 
To this aim we make two main approximations: the usual semiclassical approximation ($\hbar \ll 1$) 
and a  new one, namely the ``strongly-mixed state'' approximation.
Particular solutions of the fluid-dynamic equations are discussed which are of physical interest. 
\end{abstract}



\maketitle

\section{Introduction}
Graphene is a single layer of carbon atoms disposed as an honeycomb lattice, that is, a single sheet of graphite. 
This remarkable material has recently attracted the attention of physicists and engineers because of its interesting 
electronic properties, which make it a candidate for the construction of new electronic devices \cite{Freitag08}.
\par
Graphene is a zero-gap semiconductor, that is, the valence band of the energy spectrum 
intesects the conduction band in some points, named {\em Dirac points}; moreover, around such points the energy 
of electrons is approximately linear with respect to the modulus of momentum. 
More precisely, the Hamiltonian of an electron in a graphene lattice (which is essentially a two-dimensional system),
for low energies and in absence of external potentials is:\footnote{%
We recall the Pauli matrices:
$$\sigma_0 = \begin{pmatrix} 1 & 0\\ 0 & 1\end{pmatrix}, \quad
\sigma_1 = \begin{pmatrix}  0 & 1\\ 1 & 0 \end{pmatrix}, \quad
\sigma_2 = \begin{pmatrix}  0 & -i\\ i & 0 \end{pmatrix}, \quad
\sigma_3 = \begin{pmatrix}  1 & 0\\ 0 & -1\end{pmatrix}.$$
}
\begin{equation}
\label{H}
H_0 = -i\hbar v_F \sigma \cdot{\nabla} = -i\hbar v_F (\sigma_1\pt_x + \sigma_2\pt_y),
\end{equation}
where 
$$
  \sigma = (\sigma_1,\sigma_2), \qquad \nabla = (\pt_x,\pt_y) = \Big(\frac{\pt}{\pt x_1}, \frac{\pt}{\pt x_2}\Big).
$$
Moreover, $v_F \approx c/300 \approx 10^6\, \textrm{m/s}$, is the Fermi speed and, as usual, 
$\hbar$ denotes the  reduced Planck constant.
The corresponding energy spectrum is:
$$ 
  E = v_F \left| p \right| 
$$
with $p = (p_1,p_2)$,
which means that the electrons in graphene behave as massless relativistic particles \cite{Thaller92} 
with an ``effective light speed'' equal to $v_F$.
This remarkable feature allows to test on graphene some of the predictions of relativistic quantum mechanics
with experiments involving non-relativistic velocities.
In particular, much attention has been devoted to the so-called {\em Klein paradox}, that is, 
unimpeded penetration of relativistic particles through high potential barriers.
Let us consider a graphene sheet to which an electrostatic potential is superimposed with the shape
of a potential barrier along the direction $x$:
\begin{equation}
\label{barr}
V(x) = \left
\{\begin{array}{ll}
   V_0, & \quad a<x<b,
 \\
  0, & \quad \textrm{otherwise.}
\end{array}\right.
\end{equation}
Let us consider then a wave packet which hits such a barrier with an angle $\phi$ with respect to the
$x$ direction, and suppose that the 
amplitude of the barrier is much greater than the electron energy: $V_0 \gg E$. 
In these conditions, what actually  happens is that the electron trasmission probability is not zero at all,
but it is a quantity that is only weakly dependent from the barrier amplitude and is approximately given by
$$
 T = \frac{\cos^2\phi}{1-\cos^2(q)\sin^2\phi},
$$
where $q$ is approximately linear in $V_0$ and $\phi$-independent \cite{KatsnelsonEtAl06}.
This means that for angles close to the incident normal ($\phi\approx 0$) the trasmission probability is 
practically 1, that is, the barrier is perfectly trasparent: 
the electron penetrates unimpeded throught the barrier. 
\par
\medskip
The aim of the present paper is to deduce a fluid-dynamic model for electron transport in graphene.
Quantum fluid-dynamics is a fast-developing research field in applied mathematics, especially because
of its interest in nanoelectronics \cite{Jungel09}.
It has been boosted by the quantum formulation of the minimum entropy principle 
\cite{DegondRinghofer03,DegondMehatsRingh05}, whose application to spinorial system is very recent \cite{JSP10,Mehats09}. 
The strategy, generally speaking, is the following. 
One starts from a quantum kinetic description of the system, usually formulated in terms of Wigner
functions \cite{ZachosEtAl05}, that become matrix-valued function for spinorial systems \cite{JSP10}. 
The moments of the Wigner function are the macroscopic (fluid-dynamic) quantities of interest.
Then, fluid-dynamic equations are deduced by taking the moments of the Wigner equation (i.e.\
the evolution equation for the Wigner function).
However, in exactly the same way as for the classical Boltzmann distribution, the resulting moment 
equations are non closed, i.e.\ they contain higher-order moments. 
Then, the moment equations need to be closed, and the closure relies on the physical assumption
that the Wigner function relaxes towards a suitable equilibrium state which depends only on the
moments of interest. Such a equilibrium state is reasonably assumed to be the minimizer of a 
suitable entropy functional under the constraint of given moments.
\par
When trying to follow such strategy for electrons near a Dirac point in graphene (that is, 
for particles with Hamiltonian \eqref{H}), several difficulties arise.
First of all, one has to choose a set of moments (instead, such a choice is
standard for non-spinorial systems).
In our case, we decided to use four copies of the hydrodynamic moments (density and two components of 
the current), one for each Pauli component of the Wigner matrix, for a total of twelve moments. 
This choice is certainly not optimal for pure-states, for which it can be shown that just six of such moments
yield a closed system (see Section \ref{kfd2}).
However, as we shall see, we are interested in a regime where the mixed states are, so to speak,
``far from pure'' or {\em strongly mixed} (see Definition \ref{casomesc}), for which the twelve moments
arise more naturally from the Wigner equation.
\par
A second, deeper, difficulty comes from the fact that the Hamiltonian \eqref{H} 
is not bounded from below.
The availability of lower and lower energy states prevents the entropy functional from having a minimizer.
Probably, such a difficulty can be completely overcome only in a Fermi-Dirac entropy setting.
However, since we work with Boltzmann entropy which allows us to solve the minimization problem 
explicitly (at least in the semiclassical and strongly mixing approximations), we adopt here 
another strategy, namely the modification of the Hamiltonian with the addition of a quadratic term,
which can be physically motivated since the Dirac-point Hamiltonian is just a local approximation.
\par
\medskip
The rest  of the paper is organized as follows. 
In Section \ref{kfd} we set up the Wigner formalism and write down the Wigner equations 
of the system.
From these we deduce the non-closed system of equations for the hydrodynamic moments
and briefly describe the pure-state case.
In Section \ref{mep} we choose a quantum entropy functional and study the corresponding 
constrained minimization problem.
By making the semiclassical approximation we find an explicit solution of the minimization 
problem, as a function of the Lagrange multipliers.
Then in Section \ref{fdm}, by making the further approximation of strongly mixed states, 
the Lagrange multipliers are explicitly written as functions of the hydrodynamic moments,
which allows to close the moment system and obtain the sought hydrodynamic equations.
Finally, in Subsection \ref{fdm2}, we will focus our attention on particular solutions
of the hydrodynamic equations in the one-dimensional case, namely 
piecewise-regular and piecewise-constant solutions, which can reproduce
the Klein paradox phenomenon in this special case.
\section{Kinetic and fluid-dynamic descriptions}
\label{kfd}
\subsection{Kinetic description}
\label{kfd1}
Let us consider the quantum Liouville (or von Neumann) equation:
\begin{equation}\label{liouv}
i\hbar\partial_t S = [H,S]\,,
\end{equation}
where $S$ is the (time dependent) density operator which represents the mixed state of the system, 
and $H = H_0 + V$, where $H_0$ is given by \eqref{H} and  $V$ is an applied potential 
(e.g., the potential barrier \eqref{barr}).
The quantum-kinetic (phase-space) description of the system is based on the Wigner formulation of
quantum mechanics \cite{ZachosEtAl05}.
To explain this, let us first consider a scalar (non spinorial) density operator $S$ and let
 $\rho^S$ be its formal kernel.
The Wigner function $w = \cW\rho^S = \Op^{-1}(S)$ associated to $S$ is a function  
on phase-space defined by
\begin{equation}\label{W}
w(r,p) := (2\pi)^{-d} \int_{\R^d}\rho^S \Big(r + \frac{\hbar}{2}\xi, r - \frac{\hbar}{2}\xi\Big) e^{-ip\cdot\xi}d\xi.
\end{equation}
The reason for the notation $\Op^{-1}(S)$ is that the correspondence
$$
  S \mapsto \rho^S \mapsto w
$$
is the (formal) inverse of the Weyl quantization \cite{ZachosEtAl05,Folland89}, which associates to
a phase-space function $a$  an integral operator $A = \Op(a)$ whose kernel is the 
inverse Wigner transform of $a$, given by
\begin{equation}
\label{invw}
(\cW^{-1}a)(x,x') = \frac{1}{(2\pi\hbar)^d }\int_{\R^d} a\Big(\frac{x+x'}{2},p\Big) e^{i(x-x')\cdot p/\hbar}dp.
\end{equation}
The function $a$ is called the ``symbol'', or the ``classical symbol'', of $A$.
It can be proved that the Weyl-Wigner correspondence associates real symbols with self-adjoint operators.
The importance of Wigner transforms is evident from the following central result \cite{ZachosEtAl05,Folland89}. 
\begin{theorem} 
Let $S$ be a density operator and $w = \Op^{-1}(S)$ the associated Wigner function.
Moreover, let $a$ be a classical symbol and $A = \Op(a)$ its Weyl quantization. 
If $S A$ has finite trace, then
\begin{equation}\label{ew}
\Tr(S A) = \int_{\R^d\times\R^d}a(r,p)\, w(r,p)\,dr\,dp\,.
\end{equation}
\end{theorem}
The previous theorem states that a Wigner function $w$ behaves like a pseudo-distribution in the phase space, 
that is, it plays the role of statistical weight in observables mean computation, as the Boltzmann distribution.
However, contrarily  to the latter, $w$ is not necessarly nonnegative.
\par
\smallskip
In our spinorial case, $S$ is a $2\times 2$ matrix of operators and so is its kernel $\rho^S$.
The $2\times 2$ Wigner matrix $w$, therefore, can be defined component-wise by
$$
   w_{ij} := \cW \rho^S_{ij} = \Op^{-1}(S_{ij})
$$
where $\cW$ is defined by \eqref{W} with 
$$
  r = (r_1,r_2),\qquad p = (p_1,p_2)
 $$ 
 (recall that $d=2$ in our planar case).
It turns out that $w$ is point-wise hermitian, that is
$$
   \overline{w_{ij}(r,p)} = w_{ji}(r,p),\  \text{for all $(r,p) \in \R^2\times\R^2$.}
$$
The spinorial version of \eqref{ew} reads as follows \cite{JSP10}:
\begin{equation}
\label{ew_spin}
\Tr(SA) 
= \tr \int_{\R^2\times\R^2}a(r,p)\, w(r,p)\,dr\,dp
= \sum_{s=0}^3 \int_{\R^2\times\R^2}a_s(r,p)\, w_s(r,p)\,dr\,dp\,,
\end{equation}
where $A = \Op(a)$ is the (componentwise) Weyl quantization of the $2\times 2 $ matrix symbol $a$ 
and $w$ is the Wigner matrix of $S$. 
Here, $a_s$ and $w_s$ denote the (real) Pauli components of the matrices $\gamma$ and $w$, 
which, for any complex $2\times 2$ matrix $c$, are given by
\begin{equation}
\label{PCD}
  c_s := \frac{1}{2}\tr(\sigma_s c),  \qquad s=0,1,2,3.
\end{equation}
Note that we are using $\Tr$ for the operator trace and $\tr$ for the matrix trace.
\par
\smallskip
By applying the Wigner trasform to the von Neumann equation \eqref{liouv}, after some algebra, we obtain a 
{\em Wigner equation}, that is the evolution equation for the Wigner matrix $w$:
\begin{equation}\label{w}
\partd{w}{t} + v_F \Big(\frac{\nabla_r}{2} \cdot [\sigma, w]_+ +\frac{ip}{\hbar}\cdot[\sigma, w]\Big) + 
\Theta(V)w = 0\,,
\end{equation}
where
$$
[\sigma, w]_+ = (\sigma_1 w + w\sigma_1, \sigma_2 w + w\sigma_2)\,,\quad 
[\sigma, w] = (\sigma_1 w - w\sigma_1, \sigma_2 w - w\sigma_2)\,,
$$
and
\begin{equation}\label{theta}
(\Theta(V)w)(r,p) = \frac{i}{\hbar}(2\pi)^{-2}\int_{\R^2\times\R^2} \delta V(r,\xi)w(r,p')
e^{-i(p-p')\cdot\xi} d\xi dp'\,,
\end{equation}
with
\begin{equation*} 
\delta V(r,\xi) = V\bigg(r + \frac{\hbar}{2}\xi\bigg) - V\bigg(r - \frac{\hbar}{2}\xi\bigg)\,.
\end{equation*}
Eq.~\eqref{w} can be viewed as the analogous of the Boltzmann transport equation for our quantum spinorial system.
As already remarked, $w$ is a complex hermitian $2\times 2$ matrix and so it can be written in the Pauli  basis by means 
of four real components $w_0$, $w_1$, $w_2$ and $w_3$.
In order to shorten notations, it will be convenient to introduce the following conventions.
\begin{notation}
\label{not1}
We denote by upper indices the components of ``cartesian'' vectors, whose third components is always 
set to 0; we denote by lower indices  the components of ``spinorial'' vectors,  with three non-necessarily 
zero components.
Thus, for example, 
\begin{equation}
\label{notation}
  p = (p^1,p^2,0), \qquad \pt = (\pt^1,\pt^2,0),
\end{equation}
where
$$
  \pt^i := \frac{\pt}{\pt r_i},   \quad i = 1,2, \quad \text{and} \quad  \pt^3 := 0.
$$
Moreover, we adopt the Einstein summation convention on repeated indices.
\end{notation}
With the above conventions, Eq.~\eqref{w} in Pauli components reads as follows:  
\begin{equation}\label{wpc}
\left\{\begin{aligned}
&\partial_t w_0 + v_F \pt^j w_j + \Theta(V)w_0 = 0 
\\
&\partial_{t}w_s + v_F\Big[ \pt^s w_0 - \frac{2}{\hbar} \eta_{skj}p^k w_j\Big] + \Theta(V)w_s = 0,
\qquad s=1,2,3
\end{aligned}\right.
\end{equation}
where $\pt_t = \frac{\pt}{\pt t}$ and $\eta_{skj}$ denotes the only antisymmetric $3\times 3$ tensor which 
is invariant for cyclic permutations 
of indices and such that $\eta_{123} = 1$ (in other words,  $\eta_{skj} a_k b_j = (a \times b)_s$). 
\subsection{Fluid-dynamic description}
\label{kfd2}
Following the classical procedure, we are going to take moments of system \eqref{wpc} in order to obtain 
a set of fluid-dynamic equations for macroscopic averages. 
Let us consider the fluid-dynamic moments (spinorial densities and currents):
\begin{equation}\label{mom}
n_s(r,t) := \int w_s(r,p,t) dp\,,\quad J^k_s(r,t) := \int p^k w_s(r,p,t) dp\,,
\end{equation}
for $k=1,2$ and $s=0,1,2,3$ (recall the notations introduced in Notation \ref{not1}).
We shall also consider the spinorial velocity field $u_s$ given by
$$
 u_s^k := \frac{J_s^k}{n_s} .
$$
The moments \eqref{mom} will be the unknowns of the fluid-dynamic system. 
By multiplying \eqref{wpc} by the fluid-dynamic monomials $\{1,p^1,p^2\}$ and integrating with respect to 
$p$ in $\R^2$, we obtain:
\begin{equation}\label{fluidgen0}
\left\{\begin{aligned}
&\pt_t n_0 + v_F \pt^k n_k = 0 \\
&\pt_t n_s + v_F\Big[ \pt^s n_0 - \frac{2}{\hbar} \eta_{skj}J_j^k \Big] = 0,
\qquad s=1,2,3
\end{aligned}\right.
\end{equation}
and, for $i = 1,2$,
\begin{equation}\label{fluidgen1}
\left\{\begin{aligned}
&\pt_t J_0^i + v_F \pt^k J_k^i + n_0 \pt^i V = 0 
\\
&\pt_t J_s^i + v_F\Big[ \pt^s J_0^i - \frac{2}{\hbar} 
\eta_{s k j}\int p^i p^k w_j dp\Big] + n_s \pt^i V = 0,
\quad s=1,2,3.
\end{aligned}\right.
\end{equation}
Notice that Eqs.~\eqref{fluidgen0}--\eqref{fluidgen1} are not closed, because 
they contain the higher-moment terms
\begin{equation}\label{Qiks}
Q_s^{ik} = \int p^i p^k w_s dp,
\end{equation}
which are not writable as functions of the moments $n_s$ and $J_s^i$ without further assumptions. 
\par
\smallskip
A case where \eqref{fluidgen0}--\eqref{fluidgen1} turns out to be a (formally) closed system
is that of pure (i.e.\ non statistical) states.
Indeed, in Ref.~\cite{Bialynicki95} it is shown that, starting from the spinorial
identities
$$
  \rho  \left(\nabla_{x}  \rho\right) = 2\left(\nabla_{x} \rho_0\right) \rho,
  \qquad
   \left(\nabla_{x'}  \rho\right) \rho = 2\left(\nabla_{x'} \rho_0\right) \rho,
$$
that hold for  a pure-state density matrix 
$\rho_{ij}(x,x') = \psi_i(x) \overline{\psi_j(x')}$ ($\rho_0$ being the first Pauli component),
it is possible to deduce the relations
\begin{equation*}
\begin{aligned}
& n_s J_s^k = n_0 J_0^k,
\\
& 2\eta_{sij} n_i J_j^k = n_0 \pt^k n_s - n_s \pt^k n_0.
\end{aligned}
\end{equation*}
These equations determine the parts of the tensor $J_s^k$ that are, respectively, parallel and
orthogonal to $\vec n = (n_1,n_2,n_3)$ and one can finally deduce the formula
\begin{equation}
\label{psidentity}
 J_s^k = n_s J_0^k - \frac{1}{2} \eta_{sij}\, {n_i}\, {\pt^k} \frac{n_j}{n_0}, \quad s = 1,2,3.
\end{equation}
Hence, we see that, in the case of pure states, in the hydrodynamic  system
\eqref{fluidgen0}--\eqref{fluidgen1} the six equations 
for $n_0$,  $n_s$ and $J_0^k$ ($s = 1,2,3$, $k = 1,2$) are independent, and form a closed system 
under the closure condition \eqref{psidentity}.
\par
\smallskip
In general, i.e.\ for mixed states, the identity \eqref{psidentity} does not hold and the tensor $J_s^k$
is independent from the other moments.
In Section \ref{mep} we shall discuss a strategy for the closure of system 
\eqref{fluidgen0}--\eqref{fluidgen1} for mixed states, at least in certain physical regimes. 
\section{Minimum entropy principle}
\label{mep}
\subsection{The constrained minimization problem}
\label{mep1}
We will close the sistem \eqref{fluidgen0}--\eqref{fluidgen1}, by choosing the Wigner function as a local 
termodynamic equilibrium defined as the constrained minimizer of a suitable quantum entropy 
\cite{DegondRinghofer03,DegondMehatsRingh05}. 
We begin by introducing the \emph{quantum entropy functional}:
\begin{equation}\label{entr}
\cS(w) = \int_{\R^2\times\R^2}\tr\Big[ w\Big( \Log(w) + \frac{h}{\theta} \Big) \Big]\,dx\,dp
\end{equation}
where:
\begin{equation}\label{Log}
\Log(w) := \Op^{-1}(\log\Op(w)) 
\end{equation}
and
\begin{equation}\label{h}
h(p) = \frac{\abs{p}^2}{2m}\sigma_0 + v_F \sigma \cdot p, \qquad \sigma = (\sigma_1, \sigma_2, \sigma_3).
\end{equation}
According to \eqref{ew_spin}, $\cS(w)$ is the expected value of the quantum observable
$$
  \log S + H/\theta,
$$
where $S$ is the density operator with Wigner matrix $w$, $H$ is the Hamiltonian with 
symbol $h$ and $\theta$ is a fixed (constant) temperature.
Thus, the functional $\cS$ is not properly the entropy but, rather, it is proportional to
the Gibbs free energy
$$
  \theta \cS = \cE - \theta \cS_0,
$$
where
$$
\begin{aligned}
&\cE = \int_{\R^2\times\R^2}\tr(w h)\,dx\,dp&\qquad&\text{(energy),}
\\
&\cS_0 = -\int_{\R^2\times\R^2}\tr(w\Log(w))\,dx\,dp&\qquad&\text{(entropy).}
\end{aligned}
$$
Notice, moreover, that $h$ is the graphene Hamiltonian added with a standard,
quadratic, kinetic energy term (which we assume to be a valid approximation far from the Dirac point
$p = 0$).
In this way the Hamiltonian is bounded from below; without such term, as we shall see here below, 
the minimizer of $\cS$ would be not a summable distribution function.
\\
We can now state the constrained entropy minimization problem.
\begin{problem}
\label{minv}
Determine $w$ which minimizes the functional $\cS(w)$ given by \eqref{entr} and such that
\begin{equation}
\label{vinc}
\int_{\R^2} w_s(r,p)\,dp = n_s(r)\,,
\quad
\int_{\R^2} p^k\,w_s(r,p)\,dp = J_s^k(r)\,,
\end{equation}
for $k = 1,2$, $s = 0,1,2,3$, with $n_s$, $J_s^k$  given functions (smooth enough).
\end{problem}
In next subsection we shall deduce a necessary condition for $w$ to be solution of Problem \ref{minv}.
\subsection{Form of the constrained entropy minimizer}
\label{mep2}
First of all, we look for a more tractable form for the functional $\cS$. 
By using \eqref{ew_spin} and \eqref{PCD} we obtain:
\begin{equation}
\label{S2}
\frac{1}{2}\cS(w) 
= \sum_{s=0}^3 \int w_s \bk{\sigma_s, \Log(w)} dx\,dp + \frac{1}{\theta}\int \Big(\frac{\abs{p}^2}{2m}w_0
 + v_F\, p\cdot \vec{w} \Big) dx\,dp\,,
\end{equation}
where $\vec{w} = (w_1,w_2,w_3)$.
Then, we make the first main approximations, that is, the \emph{semiclassical approximation}
\begin{equation}
\label{semicl}
  \Log(w) = \log(w) + O(\hbar),
\end{equation}
that holds whenever $w$ does not depend on $\hbar$. 
Note that the leading term is a matrix logarithm and that in the present, spinorial, case we cannot 
exclude the presence of a non-vanishing term of 
order $\hbar$ (see e.\ g.\ Ref.~\cite{JSP10}) while in the scalar case we always have 
$\Log(w) = \log(w) + O(\hbar^2)$ \cite{DegondMehatsRingh05}.
By using the semiclassical approximation \eqref{semicl}, the properties of the Pauli matrices and the 
Taylor expansion of the logarithm, it is possible to prove the following result \cite{Zamponi_tesi}.
\begin{proposition}
\label{Sappr}
Under the assumption 
\begin{equation}\label{hplbd}
w_1^2+w_2^2+w_3^2 < w_0^2,
\end{equation}
we can write $\cS = \frac{1}{2} \widetilde{\cS} + O(\hbar)$, where 
\begin{equation}
\label{St}
\widetilde{\cS}(w) = \int \left[ w_0\left(\log(w_0) + c\left(\frac{|\vec{w}|}{w_0}\right)+ 
\frac{\abs{p}^2}{2m\theta}\right)
 + \frac{v_F}{\theta} p \cdot\vec{w}\right] dx\,dp\,,
\end{equation}
and
\begin{equation}
\label{cDef}
c(\lambda) := \frac{1}{2} \log(1-\lambda^2) + \frac{\lambda}{2}\log\left( \frac{1+\lambda}{1 - \lambda}\right).
\end{equation}
\end{proposition}
As we shall see better later on, the assumption  \eqref{hplbd}
implies that $w$ must be a mixed-state Wigner matrix. 
\par
\medskip
We can now formally solve the constrained entropy minimization problem, Problem \ref{minv}, 
with $\cS$ is replaced by its semiclassical approximation $\widetilde{\cS}$. 
For $s=0,1,2,3$, $k = 1,2$, let us define the maps $w \mapsto \hat{n}_s[w]$ and  $w \mapsto \hat{J}_s^k[w]$ 
acting on Wigner matrices $w$ and defined, 
as functions of $r$, by
$$
\hat{n}_s[w](r) = \int w_s(r,p)\,dp\,,\quad 
\hat{J}_s^k[w](r) = \int p^k w_s(r,p)\,dp\,.
$$
Let us then consider the following Lagrange multipliers problem associated 
to the constrained minimization problem.
\begin{problem}
\label{lagr}
Determine the functions $q_s^0(r)$, $q_s^k(r)$, $s=0,1,2,3$, $k = 1,2$, 
and the Wigner matrix $w$, such that
\begin{equation}\label{eqvar}
\delta\widetilde{\cS}(w) + 
\int (q_s^0\delta \hat{n}_s[w] + q_s^k\delta\hat{J}_s^k[w] )dr = 0\,.
\end{equation}
\end{problem}
By easy formal calculations we obtain
\begin{multline}
\label{dst}
\delta\widetilde{\cS}(w) = 
\int \Big\{ \Big[ \log(w_0) + c\left(\frac{|\vec{w}|}{w_0}\right)-
                     \frac{|\vec{w}|}{w_0}\, c'\left(\frac{|\vec{w}|}{w_0}\right) + \frac{\abs{p}^2}{2m\theta} + 1
  \Big]\delta w_0 
\\
 +\Big[\frac{\vec{w}}{|\vec{w}|}\, c'\left(\frac{|\vec{w}|}{w_0}\right) + 
\frac{v_F}{\theta} p \Big]\cdot\delta\vec{w}\Big\}\,dx\,dp\,,
\end{multline}
and
\begin{equation}
\label{dmom}
\delta \hat{n}_s[w] = \int \delta w_s\,dp\,,\quad 
\delta \hat{J}_s^k[w] = \int p^k\delta w_s\,dp\,.
\end{equation}
By the arbitrariness of the variations $\delta w_s$, from \eqref{eqvar} we obtain
\begin{equation}\label{eqlagr0}
\left\{\begin{aligned}
&\log(w_0) + c\left(\frac{|\vec{w}|}{w_0}\right)-
\frac{|\vec{w}|}{w_0}\,c'\left(\frac{|\vec{w}|}{w_0}\right) + q_0 = 0\,,\\
&\frac{w_s}{|\vec{w}|}\,c'\left(\frac{|\vec{w}|}{w_0}\right) + 
\frac{v_F}{\theta}p_s + q_s = 0\,,\quad s=1,2,3\,,
\end{aligned}\right.
\end{equation}
where we have defined
\begin{equation}\label{qsp}
\left\{\begin{aligned}
&q_0(r,p) := 1 + q_0^0(r) + q_0^1(r)p^1 + q_0^2(r)p^2
+ \frac{|p|^2}{2m\theta}\,,\\
&q_s(r,p) := q_s^0(r) + q_s^1(r)p^1 + q_s^2(r)p^2\,,\quad s = 1,2,3\,.
\end{aligned}\right.
\end{equation}
Equations \eqref{eqlagr0} are explicitly solvable lead straightforwardly  to 
the following result.
\begin{theorem}
\label{teoweq}
If $w^{eq}$ is the solution of Problem \ref{minv}, then there exist functions
\begin{equation}\label{qlib} 
q_s^0(r)\,,\quad q_s^k(r)\,,\qquad s=0,1,2,3\,,\quad k = 1,2\,,
\end{equation}
such that
\begin{equation}\label{weq}
\left\{\begin{aligned}
&w_0^{eq}(r,p) = \cosh(Q(r,p))e^{-q_0(r,p)}\,,
\\[3pt]
&w_s^{eq}(r,p) = q_s(r,p)\frac{\sinh(Q(r,p))}{Q(r,p)}e^{-q_0(r,p)}\,,\qquad s = 1,2,3\,,
\end{aligned}\right.
\end{equation}
where $q_0$ and $q_s$ are given by \eqref{qsp}, and 
\begin{equation}\label{qsQ}
Q := \big((q_1)^2+(q_2)^2+(q_3)^2\big)^{1/2}\,.
\end{equation}
\end{theorem}
Let us observe that, for fixed $r\in\R^2$ and for all $s = 0,1,2,3$,
$w_s^{eq}(r,\cdot)$ is a Schwartz function; in particular it is summable and all of its polynomial 
moments of arbitrary degree are finite. 
In fact
$$
  \cosh(Q(r,p)) = O(e^{c_1|p|}), \qquad q_s(r,p)\frac{\sinh(Q(r,p))}{Q(r,p)} = O(e^{c_1|p|}),
  \qquad |p|\to\infty,
$$ 
for some $c_1>0$, whereas
$$
  e^{-q_0(r,p)} = O(e^{-c_2 |p|^2}), \qquad |p|\to\infty,
$$
for a suitable $c_2>0$. 
Notice that this fact is a consequence of the choice of the free energy \eqref{entr} and of the
corrected Hamiltonian \eqref{h}.
\section{The fluid-dynamic model}
\label{fdm}
\subsection{Closure of the moment equations}
\label{fdm1}
We now come to the derivation of a fluid-dynamic model for the electron transport in graphene.
This will be achieved by closing the moment equations \eqref{fluidgen0}--\eqref{fluidgen1} with the assumption
that the system is in the local equilibrium state $w^{eq}$ given by \eqref{weq}. 
To this aim, we need to write the Lagrange multipliers 
$q_s^0(r)$, $q_s^k(r)$, $s=0,1,2,3$, $k = 1,2$, as functions of the moments \eqref{mom}. 
In order to do this, we should explicitly solve the system
\begin{equation}
\label{eqint}
\begin{aligned}
&n_0(r) = \int\cosh(Q(r,p)) e^{-q_0(r,p)} dp
\\[2pt]
&n_s(r) = \int q_s(r,p)\frac{\sinh(Q(r,p))}{Q(r,p)} e^{-q_0(r,p)} dp
\\[2pt]
&J_0^k(r) = \int p^k\cosh(Q(r,p)) e^{-q_0(r,p)} dp
\\[2pt]
&J_s^k(r) = \int p^k q_s(r,p)\frac{\sinh(Q(r,p))}{Q(r,p)} e^{-q_0(r,p)} dp
\end{aligned}
\end{equation}
with respect to the unknowns $q_s^0$ and  $q_s^k$ but, unfortunately, the integrals in \eqref{eqint} 
are not elementary solvable. 
Thus, in order to be able to perform explicit calculations we make our second main approximation.  
\begin{defi}\label{casomesc}
We say that the system is in a {\em strongly mixed state} if
\begin{equation}\label{statom}
\int Q(r,p)^2 e^{-q_0(r,p)} dp \ll \int e^{-q_0(r,p)} dp\,.
\end{equation}
\end{defi}
The reason of the name 'strongly mixed state' will be clear in the following. 
From a mathematical viewpoint, for a system in such a state, this implies that
the approximation
$$
  \int F(Q(r,p))e^{-q_0(r,p)} dp \approx \int(F(0) + F'(0)Q(r,p))e^{-q_0(r,p)} dp
$$
holds for every $F$ at least twice differentiable.
Hence, what we are assuming is that quadratic and higher terms in $Q$
are negligible in a distributional sense with respect to the statistic weight $e^{-q_0}$, 
because they do not carry a significant contribution to the computation of the integrals in \eqref{eqint}. 
So, putting 
$$
  \mu(p):= (1,p^1,p^2),
$$
we can write 
\begin{equation}\label{apprmom}
\begin{aligned}
&\int \mu(p)w_0^{eq}(r,p) {dp} \approx \int\mu(p)e^{-q_0(r,p)}dp,
\\[4pt]
&\int \mu(p)w_s^{eq} (r,p) {dp} \approx \int\mu(p)q_s(r,p)e^{-q_0(r,p)}dp,
\end{aligned}
\end{equation}
which means that we are approximating, in a distributional sense,
\begin{equation}\label{wappr}
\begin{aligned}
&w_0^{eq}(r,p) \approx \widetilde{w}_0^{eq}(r,p) := e^{-q_0(r,p)},
\\[4pt]
&w_s^{eq}(r,p) \approx \widetilde{w}_s^{eq}(r,p) := q_s(r,p)e^{-q_0(r,p)} .
\end{aligned}
\end{equation}
Under the approximations we made, i.e.\ the semiclassical and the strongly-mixing ones, 
we are able to solve equations \eqref{eqint} and write the Lagrange multipliers 
as functions of the moments.
We omit here the long but straightforward calculations and state the final result.
\begin{theorem}
\label{teoweq2}
In the assumption of strongly mixed state \eqref{statom}, the solution to Problem \ref{minv}  is given
(up to $O(\hbar)$ terms) by
\begin{equation}
\label{weq2}
w^{eq}_s = \frac{n_s}{2\pi m\theta}\bigg[1 + \frac{(u_s - u_0)\cdot(p- u_0)}{m\theta}\bigg]
\exp\bigg(-\frac{|p - u_0|^2}{2m\theta}\bigg)
\end{equation}
for $ s=0,1,2,3$. 
Such  functions are, by definition, the {\em local equilibrium Wigner distribution} of electrons in graphene.
\end{theorem}
We recall that $u_s = J_s/n_s$, for $s=0,1,2,3$, and observe that the each component $w_s^{eq}$ 
is a classical Maxwellian, with temperature parameter $\theta$, multiplied by a polynomial in $p$ of degree 1. 
In particular, 
\begin{equation}
\label{weq0}
w_0^{eq} = \frac{n_0}{2\pi m\theta}\exp\Big(-\frac{|p - u_0|^2}{2m\theta}\Big)
\end{equation}
is exactly a classical Maxwellian.
We are finally in position to perform the closure of equations \eqref{fluidgen0}--\eqref{fluidgen1} 
by assuming the system to be in the local equilibrium state described by \eqref{weq2}. 
The term \eqref{Qiks} is easily computable as a gaussian integral:
\begin{equation}
\label{Liks}
\int p^ip^kw_s^{eq}\,dp = 
n_s \left(m\theta \delta_{ik} - \frac{J_0^iJ_0^k}{n_0^2}\right)+
\frac{1}{n_0}\left(J_0^i J_s^k + J_s^iJ_0^k\right) 
=: \mathcal{L}_s^{ik}\,,
\end{equation}
for $s=1,2,3$, $i,k=1,2$. 
So, putting together \eqref{fluidgen0}, \eqref{fluidgen1}, \eqref{Liks}, we are finally able
to write the following system of quantum fluid-dynamic equations (QFDEs):
\begin{align}
\label{fluideq0}
&\left\{
\begin{aligned}
&\pt_t n_0 + v_F\pt^k n_k = 0 
\\
&\pt_t n_s + v_F\pt^s n_0 = \frac{2v_F}{\hbar} \eta_{skj}J_j^k
\end{aligned}\right.
\\[6pt]
\label{fluideq1}
&\left\{
\begin{aligned}
&\pt_t J_0^i + v_F \pt^k J_k^i = - n_0 \pt^i V 
\\
&\pt_{t} J_s^i + v_F\pt^s J_0^i = \frac{2 v_F}{\hbar} 
\eta_{skj}\mathcal{L}^{ik}_j - n_s \pt^i V 
\end{aligned}\right.
\end{align}
where $i=1,2$, $s=1,2,3$, and we used the conventions stipulated in Notation \ref{not1}.
\par
\smallskip
We end this section by explaining the meaning of the strongly-mixing assumption \eqref{statom}. 
By using \eqref{weq2} we can easily compute the integrals in \eqref{statom} and find that the 
hypothesis \eqref{statom}  is equivalent to
\begin{equation}
\label{statom2}
\frac{|\vec{n}|^2}{n_0^2} \ll \frac{1}{1 + \frac{2K}{3\theta}}\,,
\end{equation}
with $\vec{n} = (n_1,n_2,n_3)$ and
$$
 K = \sum_{j=1}^3 \frac{|u_j-u_0|^2}{2m}.
$$
In particular, since $K\geq 0$, equation \eqref{statom2} implies
\begin{equation}
\label{statom3}
\frac{|\vec{n}|^2}{n_0^2} \ll 1.
\end{equation}
Thus, if we recall that
$$
\frac{|\vec{n}|^2}{n_0^2} = 1
$$
holds  for a pure state, then we understand that we are describing states which are 
``far from pure'', that is {\em strongly mixed}.
\subsection{Particular solutions of the QFDEs}
\label{fdm2}
In this section we shall investigate some particular solutions of the QFDEs \eqref{fluideq0}--\eqref{fluideq1}. 
From now on we consider the one-dimensional case, which amounts to assuming that
all moments depends only on $r_1$ and $t$ (and not on $r_2$,) and that all components of vector moments
parallel to $r_2$-axis are identically zero. 
Thus, for the sake of brevity, we can re-define
$$
r:=r_1\,,\quad J_s = J^1_s\,,\quad u_s := u_s^1\,,\qquad s=0,1,2,3,
$$
so that  the system \eqref{fluideq0}, \eqref{fluideq1} becomes
\begin{align}
\label{fluideq1d1}
&\left\{\begin{aligned}
&\partd{n_0}{t} + v_F\partd{n_1}{r} = 0 \\
&\partd{n_1}{t} + v_F\partd{n_0}{r} = 0 
\end{aligned}\right.
\\[4pt]
\label{fluideq1d2}
&\left\{\begin{aligned}
&\partd{n_2}{t} + \frac{2 v_F}{\hbar} J_3 = 0 \\
&\partd{n_3}{t} - \frac{2 v_F}{\hbar} J_2 = 0
\end{aligned}\right.
\\[4pt]
\label{fluideq1d3}
&\left\{\begin{aligned}
&\partd{J_0}{t} + v_F\partd{J_1}{r} + n_0\totd{V}{r} = 0 \\
&\partd{J_1}{t} + v_F\partd{J_0}{r} + n_1\totd{V}{r} = 0 
\end{aligned}\right.
\\[4pt]
\label{fluideq1d4}
&\left\{\begin{aligned}
&\partd{J_2}{t} + \frac{2 v_F}{\hbar}\bigg[
m\theta\bigg(1-\frac{J_0^2}{n_0^2}\bigg)n_3 + \frac{2}{n_0} J_0 J_3 
\bigg] + n_2\totd{V}{r} = 0\\
&\partd{J_3}{t} - \frac{2 v_F}{\hbar}\bigg[
m\theta\bigg(1-\frac{J_0^2}{n_0^2}\bigg)n_2 + \frac{2}{n_0}J_0 J_2
\bigg] + n_3\totd{V}{r} = 0
\end{aligned}\right.
\end{align}
where we immediately notice that the components 0 and 1 are completely decoupled from the components 2 and 3.
\par
Let us consider the system \eqref{fluideq1d1}--\eqref{fluideq1d4} when $V$ is the potential barrier \eqref{barr}. 
In this case, the derivative of $V$, appearing in the equations, has to be intended in distributional sense:
$$
  \totd{V}{r} = V_0(\delta(r-a)-\delta(r-b))\,,
$$
where $\delta(r-a)$ is the delta distribution centered in $a$. 
The derivatives of moments, also appearing in the equations, will
be considered as distributional derivatives, too.
Let us consider the sets
$$
\begin{aligned}
  &\Omega := \Set{(r,t)\in\R^2}{r \neq a\,,\, r \neq b}, &\quad &\Omega_1 := \Set{(r,t)\in\R^2}{r<a}\,,
\\[2pt]
  &\Omega_2 := \Set{(r,t)\in\R^2}{a<r<b},&\quad &\Omega_3 := \Set{(r,t)\in\R^2}{r>b}\,,
\end{aligned}
$$
and let us define the space $X\subset L^1_{\mathrm{loc}}(\R^2)$ in the following way: 
for each $u: \R^2\to\R$ we say that $u \in X$ if and only if
\begin{enumerate}
\item $u\in C^1(\Omega)$;
\item for all $t_0 \in \R$, the limits 
$$
\begin{aligned}
& \lim_{(r,t)\to (a^-,t_0)} u(r,t), &\qquad &\lim_{(r,t)\to (a^+,t_0)}  u(r,t),
\\[2pt]
& \lim_{(r,t)\to (b^-,t_0)} u(r,t), &\qquad &\lim_{(r,t)\to (b^+,t_0)}  u(r,t),
\end{aligned}
$$
exist and are finite.
\end{enumerate}
For $u\in X$ and $(r_0,t)\in\R^2$ arbitrary, let us define then the \emph{jump of $u$ in $r_0$} as
$$
  [u]_{r_0}(t) := \lim_{r \to r_0^{+} }u(r,t) -  \lim_{r \to r_0^-}u(r,t)\,.
$$
Let us indicate with $\partial_r u$ the distributional derivative of $u$, and with $\partd{u}{r}$ the almost
everywhere derivative of $u$. 
Then, we have the following (the proof is standard).
\begin{lemma} 
If $u\in X$, then
\begin{equation}
\label{derdistr}
\partial_r u = \partd{u}{r} + [u]_a\delta(r - a) + [u]_b\delta(r - b)\,.
\end{equation}
\end{lemma}
We now consider piecewise-regular solutions of \eqref{fluideq1d1}--\eqref{fluideq1d4}, according to the
following definition.
\begin{defi}
A 8-tuple of real-valued functions $(n_s,J_s)_{s=0,1,2,3}$ defined in $\R^2$ is \emph{piecewise-regular} if
\begin{equation}
\label{ipot}
n_s\in C(\R^2)\,,\quad\partd{n_s}{r}\in X,\quad  J_s \in X\,,\qquad s = 0,1,2,3\,.
\end{equation}
\end{defi}
From the previous lemma, assuming that eqs.\ \eqref{fluideq1d1}--\eqref{fluideq1d4} have a piecewise-regular
solution $(n_s,J_s)_{s=0,1,2,3}$,  we immediately deduce that in such equations the potential terms can be 
written
\begin{equation}\label{ndv}
n_s \totd{V}{r} = \sum_{r_0 = a,b}n_s(r_0,t)[V]_{r_0}\delta(r - r_0)\,,\quad s=0,1,2,3.
\end{equation}
Let us, then, consider the first of eqs.~\eqref{fluideq1d3} written as
\begin{equation}
\label{jdelta}
\partd{J_0}{t} + v_F \partd{J_1}{r} = - \sum_{r_0 = a, b} (v_F [J_1]_{r_0} + n_0(r_0,t)[V]_{r_0})\delta(r - r_0).
\end{equation}
By integrating both sides for $r \in (a-\epsilon, a + \epsilon)$ (with $0<\epsilon< b - a$) we obtain
$$
  \int_{a-\epsilon}^{a + \epsilon} \Big(\partd{J_0}{t} + v_F \partd{J_1}{r}\Big) dr
   = -(v_F [J_1]_a + n_0(a,t) [V]_a).
$$
If $\epsilon\to 0$, then the first side tends to $0$, by the integrability of the involved functions, and so
$$
  v_F [J_1]_a(t) + n_0(a,t) [V]_a = 0.
$$
Analogously, by integrating eq.~\eqref{jdelta} in a neighborhood of $b$, we deduce
$$
  v_F [J_1]_b(t) + n_0(b,t) [V]_b = 0,
$$
and so, again from \eqref{jdelta}, we also find 
$$
  \partd{J_0}{t} + v_F \partd{J_1}{r} = 0.
$$
From the second of \eqref{fluideq1d3} we get in the same way:
$$\partd{J_1}{t} + v_F \partd{J_0}{r} = 0\,,
$$ $$
  v_F [J_0]_a(t) + n_1(a,t) [V]_a = v_F [J_0]_b(t) + n_1(b,t) [V]_b = 0.
$$
By repeating this reasoning for eq.~\eqref{fluideq1d4}, we are finally led to the following theorem.
\begin{theorem}
\label{teocs}
If $M := (n_s,J_s)_{s=0,1,2,3}$ is a piecewise-regular solution to system \eqref{fluideq1d1}--\eqref{fluideq1d4}, 
with $V$ given by \eqref{barr}, then $M$ satisfies the same equations  with $V\equiv 0$ in the set $\Omega$,
together with the following \emph{jump conditions}:
\begin{equation}
\label{salto}
\left\{\begin{aligned}
& v_F [J_1]_{r_0}(t) + n_0(r_0,t) [V]_{r_0} = 0\\
& v_F [J_0]_{r_0}(t) + n_1(r_0,t) [V]_{r_0} = 0\\ 
&n_2(r_0,t) = n_3(r_0,t) = 0
\end{aligned}\right.
\end{equation}
for $r_0 = a,b$. 
Conversely, if $M$ satisfies \eqref{fluideq1d1}--\eqref{fluideq1d4} with $V\equiv 0$ in the set $\Omega$ 
and the conditions \eqref{salto} in $r_0 = a,b$, then $M$is a piecewise-regular solution to
eqs.~\eqref{fluideq1d1}-\eqref{fluideq1d4}, with $V$ given by \eqref{barr}.
\end{theorem}
Let us observe that the first two jump conditions can be interpreted as conservation laws: 
the first condition represents conservation of energy, while the second one 
is a momentum balance. 
In particular, in the present one-dimensional case, the total energy density at $(r,t)$ is  
\begin{multline}
\label{Hm}
\bk{H}(r,t) = 
\frac{1}{n_0(r,t)}\int \left[ v_Fp_1w_1(r,p,t) + \left( \frac{1}{2m}|p|^2 + V(r) \right) w_0(r,p,t) \right] dp
\\
= v_F \frac{J_1(r,t)}{n_0(r,t)} + \frac{\theta}{2} + V(r)
\end{multline}
(where we used \eqref{weq0}), and then, since $n_0$ is continuous, the first of Eqs.~\eqref{salto}
reads as 
$$
[\bk{H}]_{r_0} = \frac{v_F}{n_0(r_0,t)}[J_1]_{r_0}(t) + [V]_{r_0} = 0.
$$
Let us now focus on a particular class of solutions: the piecewise-constant ones.
\begin{defi}
A 8-tuple of real-valued functions $M = (n_s,J_s)_{s=0,1,2,3}$ defined in $\R^2$ is 
\emph{piecewise-constant} if, for $s = 0,1,2,3$,
\begin{enumerate}
\item[a)] $n_s$ is constant with respect to $(r,t)\in\R^2$;
\item[b)] $J_s$ is constant with respect to $(r,t)\in \Omega_j$, $j = 1,2,3$.
\end{enumerate}
\end{defi}
Because such a 8-tuple of functions satisfies obviously the equations 
\eqref{fluideq1d1}-\eqref{fluideq1d4} in $\Omega$, then Theorem \ref{teocs} implies the following.
\begin{proposition}
The piecewise-constant solutions of the system \eqref{fluideq1d1}-\eqref{fluideq1d4} are
given by
\begin{equation}
\label{solcost}
\left\{\begin{aligned}
n_2 = n_3 = J_2 = J_3 = 0\,, \\
v_F J_0(r) + n_1 V(r) = \beta_0\,,\\
v_F J_1(r) + n_0 V(r) = \beta_1\,,
\end{aligned}\right.
\end{equation}
with $\beta_0, \beta_1$ constants. 
\end{proposition}
Such piecewise-constant solutions of the QFDEs are linked with the 
solutions of the Schr\"odinger equation used in \cite{KatsnelsonEtAl06} 
to describe Klein paradox in graphene, in the one-dimensional case.
Indeed, an electron wave incident the potential barrier with angle $\phi=0$ 
(that is, perpendicularly)
is described by the following components of a spinorial wave function:
\begin{equation}
\label{fonda}
\begin{aligned}
&\psi_1(x,y) = \left\{\begin{array}{ll}
e^{ikx}  & \quad x < a \\
\alpha e^{iqx} + \beta e^{-iqx} & \quad a < x < b \\
t e^{ikx} & \quad x > b
\end{array}\right.
\\[4pt]
&\psi_2(x,y) = \left\{\begin{array}{ll}
s e^{ikx}  & \quad x < a \\
s'(\alpha e^{iqx} - \beta e^{-iqx}) & \quad a < x < b \\
st e^{ikx} & \quad x > b
\end{array}\right. 
\end{aligned}
\end{equation}
where
$$
  q = \frac{|E - V_0|}{\hbar v_F}\,,\quad s = \sign E ,\quad s' = \sign (E - V_0),\quad
|E| = \hbar k v_F,
$$ $$
 t = e^{iD (\frac{s}{s'}q - k)},\quad\,\alpha = 
\frac{1}{2}\left(1 + \frac{s'}{s}\right),\quad \beta = \frac{1}{2}\left(1 - \frac{s'}{s}\right).
$$
We recall that the trasmission probability is $T = |t|^2 = 1$ and, therefore, 
we have perfect tunneling. 
If we compute the moments $n_s$, $J_s$ associated to the wave function \eqref{fonda}, we find
\begin{equation}
\label{varonde}
\left\{\begin{aligned}
&n_0 = 1,\quad n_1 = s,\quad n_2 = n_3 = 0;
\\
&J_0 = (E - V(r))\frac{n_1}{v_F},\quad J_1 = (E - V(r))\frac{n_0}{v_F}; 
\\
&J_2 = J_3 = 0.
\end{aligned} \right.
\end{equation}
Notice that there is a perfect agreement between \eqref{solcost} and \eqref{varonde}. 
In particular, we can relate the constants $\beta_0$ and $\beta_1$ to the electron energy $E$:
$$
  \beta_0 = n_1 E,\qquad \beta_1 = n_0 E.
$$
This fact suggests that the equation \eqref{fluideq0}, \eqref{fluideq1} are suitable to 
describe electronic tunneling in graphene, at least in the one-dimensional case:
the solutions \eqref{solcost} represent exactly such a phenomenon.


\end{document}